\documentclass[10pt, conference, letter]{IEEEtran}
\IEEEoverridecommandlockouts

\usepackage{amsmath, amsfonts, amssymb}
\usepackage{xcolor}
\usepackage{comment}
\usepackage{graphicx} 
\usepackage{multirow} 
\usepackage{hhline} 
\usepackage{subcaption} 
\usepackage{adjustbox} 
\usepackage{tikz}
\usepackage{nicematrix}
\usepackage{soul}

\definecolor{myblue}{HTML}{B8E0F2}
\definecolor{mygreen}{HTML}{C5E8D8}
\definecolor{mypeach}{HTML}{FFE6C7}

\DeclareRobustCommand{\cnum}[1]{%
  \tikz[baseline=(char.base)]{
    \node[draw,circle,inner sep=1pt] (char) {#1};
  }%
}

\usetikzlibrary{positioning}
\usepackage{circledtext}

\title{Low-Cost Multi-Precision Systolic Arrays for Accelerating FHE NTTs on AI ASICs} 
\author{%
\IEEEauthorblockN{George Alexakis}
\IEEEauthorblockA{Electrical and Computer Engineering\\ 
Democritus University of Thrace, Xanthi, Greece}
\and
\IEEEauthorblockN{Dimitrios Schoinianakis}
\IEEEauthorblockA{Nokia Bell Labs\\  
Athens, Greece}
\and
\IEEEauthorblockN{Giorgos Dimitrakopoulos}
\IEEEauthorblockA{Electrical and Computer Engineering\\ 
Democritus University of Thrace, Xanthi, Greece}
}

\begin{document}

\maketitle

\begin{abstract}
Fully Homomorphic Encryption (FHE) ensures robust data privacy but suffers from prohibitive computational overhead. Accelerating FHE on AI hardware like Tensor Processing Units (TPUs) is promising, yet fundamentally limited by a precision mismatch: TPUs are optimized for 8-bit arithmetic, whereas FHE
and its critical parts such as the Number Theoretic Transform (NTT), demand high precision.
Current approaches bridge this gap using matrix decomposition to execute NTT computations on low-precision matrix engines. 
However, reconstructing the full-precision results requires shift-and-add accumulation that does not match the dataflow of matrix multiplication. This forces offloading full-precision reconstruction from matrix engines to vector processors that disrupts the matrix multiplication dataflow, creating significant performance bottleneck. 
To resolve this limitation, we propose a minimally modified multi-precision systolic array that performs  full-precision output reconstruction natively within the array \emph{in sync} with low-precision matrix multiplication under a uniform dataflow. Synthesized at 7nm with OpenRoad, our design incurs negligible hardware overhead. Cycle-accurate simulations using SCALE-Sim demonstrate that natively executing NTTs on the proposed architecture achieves at least 1.33$\times$ speedup, for transform sizes $2^{12}$ to $2^{16}$ on 128$\times$128 matrix engines, successfully enabling standard AI hardware to support high-precision FHE acceleration.
%
%
%
\end{abstract}

\section{Introduction}

Fully Homomorphic Encryption (FHE) offers a powerful security guarantee: it enables arbitrary computation directly on encrypted data without ever exposing plaintext. Despite this promise, its extreme computational overhead has traditionally necessitated specialized, custom-built hardware~\cite{wang2012accelerating,
samardzic2021f1}. Designing such bespoke accelerators, however, is prohibitively expensive and difficult to scale. A compelling alternative is to leverage existing AI accelerators~\cite{10071017, 11230299, 11408507}, such as Tensor Processing Units (TPUs), which already deliver massive parallelism and high energy efficiency~\cite{9499913}. Repurposing these platforms provides a practical path toward closing the performance gap that currently limits real-world FHE adoption~\cite{samardzic2021f1}.

Mapping FHE workloads to AI accelerators reveals a fundamental architectural mismatch: TPUs are optimized for 8-bit integer arithmetic~\cite{9499913}, whereas FHE demands modular arithmetic over large integer moduli~\cite{acar2018survey}. Although FHE mitigates this using the Residue Number System~\cite{bajard2016full}, which splits large integers into smaller, independent 30-60 bit channels, these components remain too large for native TPU matrix engines.

To utilize TPU matrix engines, a common workaround is to decompose high-precision values into multiple low-precision (e.g., 8-bit) matrix operations, whose results must later be aggregated to full-precision. 
This reconstruction requires a sequence of shift-and-accumulate operations and digit-level carry propagation. Current TPU architectures cannot perform these operations within their systolic matrix engines, forcing computation to spill over to external vector units. This ``external accumulation'' incurs substantial data movement overhead, disrupts matrix multiplication dataflow, and leaves the matrix engine underutilized, effectively negating the advantages of the accelerator.

In this work, we show that efficient FHE execution on AI accelerators is achievable with minimal hardware changes. We propose a lightweight modification to the processing elements of the last row of a systolic array that enables in-place shift-and-add operations directly within the matrix engine. 
Unlike prior approaches for multi-precision matrix computation that rely on reconfigurable SIMD-style systolic arrays~\cite{camus2019survey}, our design introduces negligible hardware overhead while supporting multi-precision accumulation natively. By embedding reconstruction into the uniform dataflow of matrix multiplication, we eliminate the need for external accumulation and unlock sustained high utilization of the accelerator.

This approach bridges the gap between the high-precision requirements of FHE and the low-precision efficiency of modern AI hardware, enabling practical deployment without sacrificing compatibility with existing workloads. The contributions of this work can be summarized as follows:
\begin{itemize}
\item  We identify the full-precision reconstruction phase, specifically shift-and-add accumulation and digit-level carry propagation, as a critical performance bottleneck when mapping core FHE operations, such as the Number Theoretic Transform (NTT) used to accelerate polynomial multiplications, onto AI accelerators.
\item We introduce a minimally modified systolic array that performs full-precision reconstruction internally and in sync with low-precision matrix multiplication under a uniform dataflow. Supporting inside the systolic array this dual operation of low-precision matrix multiplication and full-precision result reconstruction adds negligible hardware cost (less than 1\%), as quantified by implementation results at 7nm using OpenROAD~\cite{ajayi2019openroad}.
\item We evaluate performance using the SCALE-Sim cycle-accurate simulator. For the NTT, which dominates FHE runtime, we demonstrate at least 1.33$\times$ speedup
for transform sizes ranging from $2^{12}$ to $2^{16}$, while preserving compatibility with conventional low-precision AI matrix engines and incurring minimal overhead.
\end{itemize}

\section{Background and Related Work}

FHE schemes are dominated by linear algebra and polynomial arithmetic~\cite{halevi2014algorithms}. Ciphertexts are represented as vectors of ring elements, and core operations such as key switching, automorphisms, and ciphertext multiplication ultimately reduce to structured linear transformations and polynomial convolutions~\cite{10071017,11230299,halevi2014algorithms,11408507}. 

Recent work has shown that many of these kernels can be reformulated using dense matrix operations, enabling acceleration on GPUs and AI accelerators. TensorFHE~\cite{10071017} expresses FHE primitives for GPU tensor cores, while TensorFHE+~\cite{11230299} extends this idea through a linear-algebra abstraction that improves portability across heterogeneous accelerators. CROSS~\cite{11408507} further demonstrates the feasibility of mapping FHE workloads onto systolic-array accelerators by reformulating high-precision modular arithmetic into low-precision matrix multiplications. This transformation is particularly critical for core FHE operations~\cite{harvey2019faster}, such as the NTT kernels that simplify polynomial multiplication.


\subsection{Matrix-Based NTT and Mapping to AI ASICs}
Custom FHE accelerators typically rely on dedicated parallel butterfly units to compute the NTT~\cite{samardzic2021f1}. In contrast, approaches that target AI ASICs leverage existing matrix multiplication engines and vector processors. In this setting, the large NTTs required by FHE applications (ranging from $2^{12}$ to $2^{16}$ elements) are executed using the 4-step NTT algorithm~\cite{bailey1990fft}, which decomposes the monolithic transform into a sequence of matrix-friendly sub-operations.

To execute the 4-step NTT algorithm on an 1D input of size $N = r \times c$ the input is first reshaped into an $r \times c$ matrix. Then, the following four steps are executed:\\
{\it Column-wise Transforms:} The algorithm performs $c$ independent $r$-point NTTs along the columns of the matrix. This is executed as a matrix multiplication between a precomputed $r \times r$ transform matrix and the $r \times c$ data matrix.\\
{\it Twiddle Factor Multiplication:} To correct for the phase interactions introduced by moving to two dimensions, the intermediate results undergo an element-wise multiplication with a precomputed $r \times c$ matrix of twiddle factors.\\
{\it Matrix Transpose:} The $r \times c$ matrix is transposed into a $c \times r$ layout. This structural reorganization aligns the original row data into contiguous columns.\\
{\it Row-wise Transforms:} Finally, $r$ independent $c$-point NTTs are executed along the newly transposed columns. This is computed via a second matrix multiplication, this time using a $c \times c$ transform matrix. 


Ultimately, the 4-step decomposition replaces a single $N$-point NTT with two $\sqrt{N} \times \sqrt{N}$ matrix multiplications, one element-wise multiplication, and one transpose operation. This formulation aligns well with the systolic matrix units found in modern AI accelerators: systolic arrays can natively execute the matrix multiplications, while auxiliary processing units handle the transpose and element-wise operations. 

\subsection{Precision mismatch between FHE and AI ASICs}
Mapping FHE workloads onto AI ASICs exposes a significant architectural mismatch. FHE requires modular arithmetic over large integer moduli (often spanning hundreds of bits), whereas modern AI ASICs rely on 8-bit integer units to maximize energy efficiency and scalability (typically $128 \times 128$ processing elements or larger)~\cite{9499913}.

To bridge this precision gap, existing approaches, such as the optimized mapping in CROSS~\cite{11408507}, decompose high-precision matrix operations into multiple low-precision sub-steps. The resulting partial outputs are then aggregated to recover the desired precision. However, this reconstruction/aggregation stage becomes a critical performance bottleneck, as it effectively serializes execution after the matrix multiplications. Moreover, because reconstruction is offloaded to vector processors outside the main matrix engine, it introduces data movement overhead and reduces systolic-array utilization.

This inefficiency motivates our work: we integrate high-precision recomposition directly into the matrix multiplication dataflow. Our goal is to augment standard 8-bit systolic arrays to also perform the reconstruction step, producing full-precision matrix multiplication results without requiring computation outside the array.

Crucially, we achieve this without relying on traditional reconfigurable SIMD-based multi-precision architectures~\cite{camus2019survey,mao2022energy}, which group adjacent processing elements to emulate wider arithmetic units. While effective in other domains, such spatial grouping incurs significant hardware overhead and reduces the effective array size available for parallel matrix multiplication as precision requirements increase.

\section{Multi-Precision Matrix Multiplication in Enhanced Systolic Arrays}

TPU matrix engines (systolic arrays) are built around small, fixed-width multipliers and relatively wide accumulators. For instance, Google TPUs contain 8-bit integer multipliers in each processing element, while the column-wise accumulation (reduction) in a weight-stationary dataflow uses 32-bit adders~\cite{9499913}.

This design makes them highly efficient for low-precision dense linear algebra. However, it also means that high-precision arithmetic, where input operands exceed 8 bits, cannot be executed directly within the array. Instead, larger integers must be decomposed into smaller digits, and matrix multiplication must be carried out over the corresponding low-precision elements. Intermediate results must then be aggregated using shift-and-add operations to reconstruct the final full-precision result.

Systolic arrays are not designed to perform full-precision reconstruction internally; instead, this step is typically offloaded to vector processors outside the array that disrupts the otherwise uniform dataflow of matrix multiplication, 
introducing unnecessary data movement and effectively lowering the utilization of the matrix engine.

To address this, we first show how full-precision input matrices can be decomposed into matrices composed of small digits (e.g., 8-bit integers). We then describe how standard matrix multiplication generates intermediate results, and finally how these results can be processed within the systolic array to recover the correct full-precision output as a seamless extension of the original dataflow.

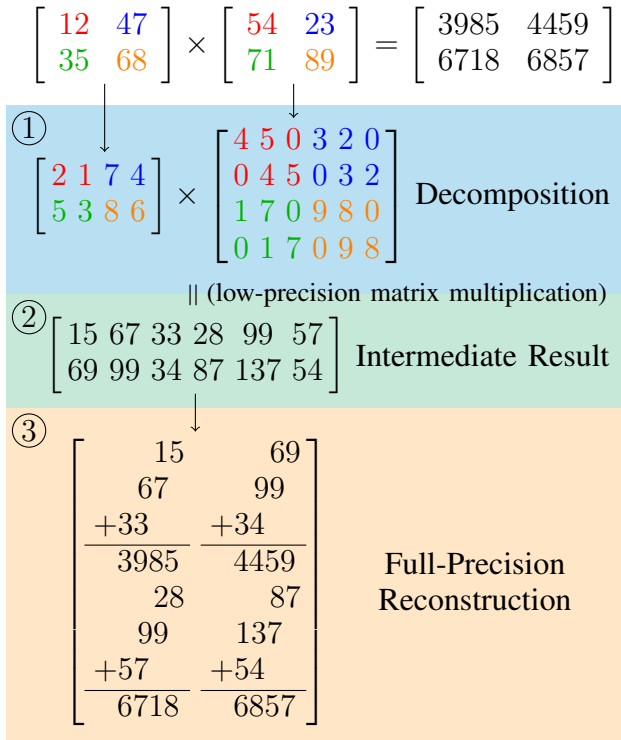
\begin{figure}[t]
    \centering
    \begin{tikzpicture}
        \fill[myblue] (-4.2,-3.3) rectangle (4,-0.8);
        \fill[mygreen] (-4.2,-4.8) rectangle (4,-3.3);
        \fill[mypeach] (-4.2,-9.2) rectangle (4,-4.8);
        \node[on grid, font=\large, inner sep=0pt] (m1) {$\left[
            \begin{array}{cc}
            \textcolor{red}{12} & \textcolor{blue}{47} \\
            \textcolor{green!70!black}{35} & \textcolor{orange}{68} \\
            \end{array}\right]
            \times
            \left[
            \begin{array}{cc}
            \textcolor{red}{54} & \textcolor{blue}{23} \\
            \textcolor{green!70!black}{71} & \textcolor{orange}{89} \\
            \end{array}\right]
            =
            \left[
            \begin{array}{cc}
            3985 & 4459 \\
            6718 & 6857 \\
            \end{array}\right]$};
        
        \node[on grid, font=\large, inner sep=0pt, below left = 2cm and 1.4cm of m1] (m2) {\setlength{\arraycolsep}{2pt}$\left[
            \begin{array}{cccc}
            \textcolor{red}{2} & \textcolor{red}{1}   & \textcolor{blue}{7} & \textcolor{blue}{4} \\
            \textcolor{green!70!black}{5} & \textcolor{green!70!black}{3} & \textcolor{orange}{8} & \textcolor{orange}{6} \\
            \end{array}\right]
            \times 
            \left[
            \begin{array}{cccccc}
\textcolor{red}4 & \textcolor{red}5 & \textcolor{red}0 & \textcolor{blue}3 & \textcolor{blue}2 & \textcolor{blue}0 \\
\textcolor{red}0 & \textcolor{red}4 & \textcolor{red}5 & \textcolor{blue}0 & \textcolor{blue}3 & \textcolor{blue}2 \\
\textcolor{green!70!black}1 & \textcolor{green!70!black}7 & \textcolor{green!70!black}0 & \textcolor{orange}9 & \textcolor{orange}8 & \textcolor{orange}0 \\
\textcolor{green!70!black}0 & \textcolor{green!70!black}1 & \textcolor{green!70!black}7 & \textcolor{orange}0 & \textcolor{orange}9 & \textcolor{orange}8 \\
            \end{array}\right]$};

        \node[on grid, inner sep=0pt, font=\large, below left = 2.1cm and 0.3cm of m2] (m3) {\setlength{\arraycolsep}{2pt}$\left[
            \begin{array}{cccccc}
            15 & 67 & 33 & 28 & 99 & 57 \\
            69 & 99 & 34 & 87 & 137 & 54 \\
            \end{array}\right]$};
            
        \node[on grid, inner sep=0pt, font=\large, below = 3cm of m3] (m4) {\setlength{\arraycolsep}{2pt}$\left[
            \begin{array}{ccc}
            \quad\quad15 && \quad\quad69 \\
            \quad67 && \quad99 \\
            +33\quad && +34\quad \\
            \hhline{-~-}
            \ \ 3985 && \ \ 4459 \\
            \quad\quad28 && \quad\quad87 \\
            \quad99 && \ 137  \\
            +57\quad && +54\quad \\
            \hhline{-~-}
            \ \ 6718 && \ \ 6857 \\
            \end{array}\right]$};

\node[on grid, rotate=90, inner sep=0pt, font=\large, above = 0.8cm of m3] (eq) {$=$};
\node[on grid, inner sep=0pt, font=\large, right = 3.9cm of m2] (d) {Decomposition};
\node[on grid, inner sep=0pt, font=\large, right = 3.8cm of m3] (i) {Intermediate Result};
\node[on grid,align=center, inner sep=0pt, font=\large, right = 3.7cm of m4] (r) {Full-Precision \\  Reconstruction};

\node[on grid, circle, draw, inner sep=1pt, font=\large, above left = 0.9cm and 2.5cm of m2] (c1) {$1$};
\node[on grid, circle, draw, inner sep=1pt, font=\large, below = 2.5cm of c1] (c2) {$2$};
\node[on grid, circle, draw, inner sep=1pt, font=\large, below = 1.5cm of c2] (c3) {$3$};
\node[on grid, inner sep=1pt, right  = 2.8cm of eq] (c4) {(low-precision matrix multiplication)};
            \draw[->] (m3.south) -- ++(0,-0.5);
            \draw[->] ([xshift=-0.4cm]m1.south) -- ++(0,-0.45);
            \draw[->] ([xshift=-2.9cm]m1.south) -- ++(0,-0.9);
    \end{tikzpicture}
    \caption{Conversion of high precision matrix multiplication to low precision, through decomposition and reconstruction.}
    \label{fig:highgemm}
\end{figure}


\subsection{Overview of the High-Precision Matrix Multiplication Flow}
\label{sec:decomp}

Fig.~\ref{fig:highgemm} illustrates how a matrix multiplication with multi-digit (high-precision) integers is transformed into an equivalent matrix multiplication over single-digit operands that can be computed efficiently on a systolic array. The example uses two-digit decimal numbers for clarity, but the same procedure applies to any other radix.

The top of the figure shows the result of the multiplication of the full-precision matrices. 
%
%
Each full-precision matrix is first decomposed into an larger matrix consisting solely on smaller digits (step \cnum{1} in Fig.~\ref{fig:highgemm}). For the left matrix, every element is expanded into a row vector containing the original digits but placed to consecutive columns from the least significant to most significant digit. This produces a wider matrix whose entries are single-digit values. 
The right matrix undergoes a different transformation. Instead of simply stacking digits, the constituent digits are arranged into a Toeplitz-like structure~\cite{gray2006toeplitz} that encodes all digit cross-products required by schoolbook digit-wise multiplication~\cite{knuth2014art}. Each digit of the right-hand matrix is copied into multiple columns with appropriate horizontal shifts. These shifts ensure that when the single-digit matrices are multiplied, products between digit pairs automatically align with the correct positional weight. As a result, a single dense matrix multiplication generates all digit cross-products required by the original high-precision computation. 

To clarify this, let's assume two entries of the corresponding input matrices, $a$ is an element of left matrix and $b$ an element of the write matrix. Since both contain two decimal digits we can write 
$a = a_1 \cdot 10 + a_0$ and $b = b_1 \cdot 10 + b_0$, where $a_1, a_0$ and $b_1, b_0$ represent each digit. 
When we multiply them ($a \times b$), you get a quadratic polynomial for:
$$c = (a_1 b_1) \cdot 10^2 + (a_1 b_0 + a_0 b_1) \cdot 10^1 + (a_0 b_0) \cdot 10^0$$
We can express the calculation of these three coefficients as a matrix-vector multiplication
\begin{equation*}
\begin{bmatrix}
b_0 & b_1
\end{bmatrix}
\begin{bmatrix}
a_0 & a_1 & 0 \\
0   & a_0 & 0
\end{bmatrix} = 
\begin{bmatrix}
a_0b_0 & a_1 b_0 + a_0 b_1 & a_1b_1
\end{bmatrix}
\end{equation*}
In this example and in the one shown in Fig.~\ref{fig:highgemm}, the decomposed digits are intentionally placed least-significant digit first. This property is essential for proper operation scheduling within the systolic array, as explained in the next section.

The multiplication of the single-digit matrices yields a $2\times 6$ matrix of intermediate results (step \cnum{2} in Fig.~\ref{fig:highgemm}).
Recomposition then is required to combine the intermediate results to form the correct full-precision result.
As shown in part \cnum{3} of Fig.~\ref{fig:highgemm} to perform this reconstruction step, neighbor columns corresponding to different digit positions should be combined through shifts and digit-level carry propagation thus completing the multi-digit addition. Carry propagation remains within the limits of each group of digits and does not expand to columns of different output digits.

\subsection{Proposed Systolic Array Architecture}


Performing precision reconstruction natively within systolic arrays without disrupting their typical dataflow requires almost no changes to their standard operation. We only need to augment the PEs of the final row of the systolic array with an additional adder, while the rest of the array remains unmodified.

\begin{figure}[h]
    \centering
    \includegraphics[width=0.95\columnwidth]{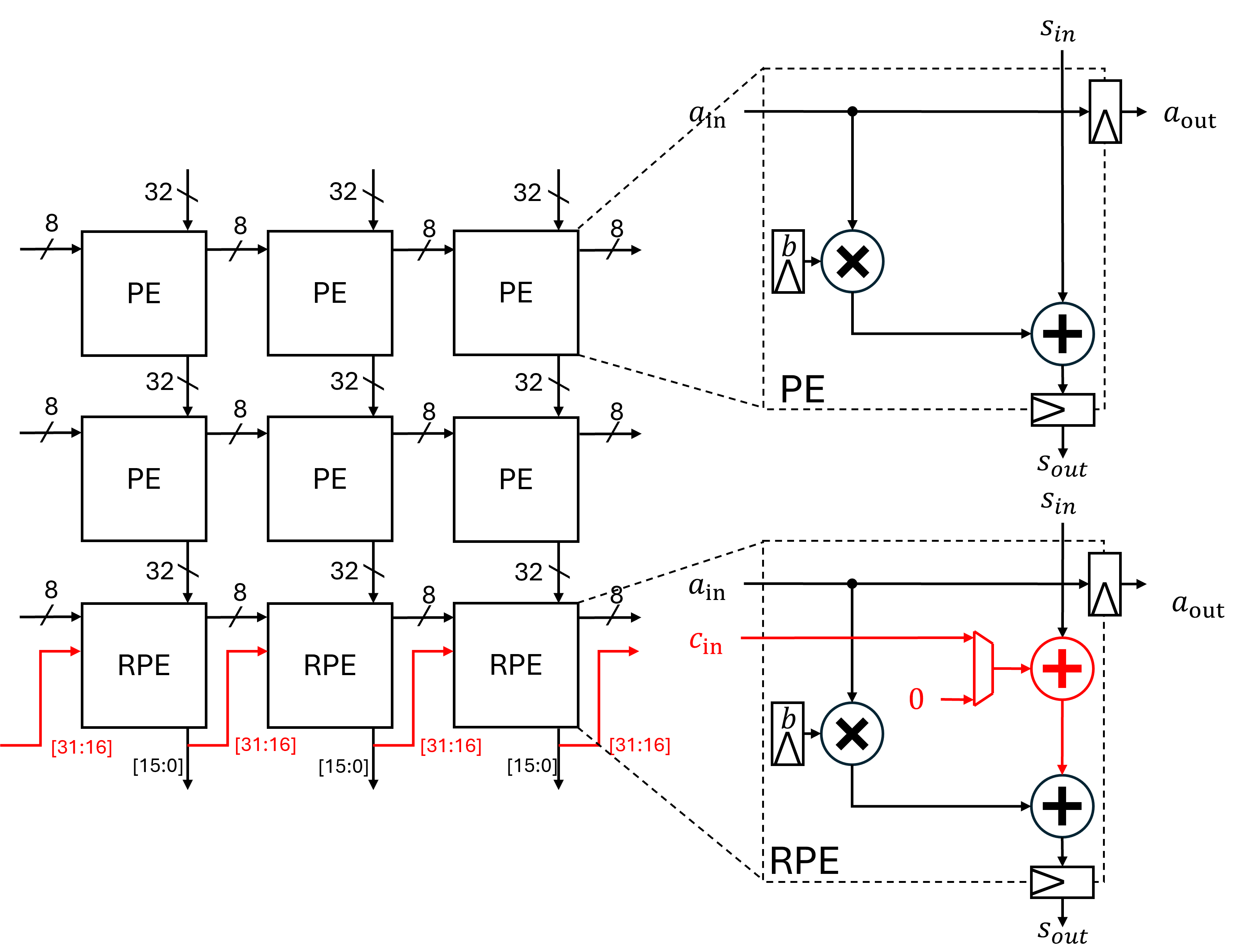}
    \caption{The proposed systolic array. All PEs keep their typical multiply-add structure while the PEs of the last row (RPEs) are enhanced with an additional adder that combines horizontal digit-level carry propagation with vertical per-column addition.}
    \label{fig:array}
\end{figure}

We base our architecture on a standard Google TPU-style systolic array, utilizing the widely adopted weight-stationary dataflow. In this baseline design, the right matrix (weights) is preloaded into the array, while the left matrix streams horizontally into the array from the western edge. Each PE performs an 8-bit multiplication between the incoming streaming data and its locally stored weight. The 16-bit products are then continuously accumulated vertically down each column using 32-bit adders. This typical structure is depicted in Fig.~\ref{fig:array}.

The PEs of the last row called Reconstruction PEs (RPEs) include one additional input relative to the typical PE of the rest rows. These RPEs receive a digit from the RPE on the left and add it to the sum propagating downwards on the same column. In this way, digit carries are propagated horizontally along the final row of the array in sync with the vertical propagation of the per-column sums. The multiplexers stops this carry propagation by selecting a zero input to the adder when result exceeds the columns of the full-precision result.


\begin{figure}[t]
    \centering
    \includegraphics[height=0.78\textheight]{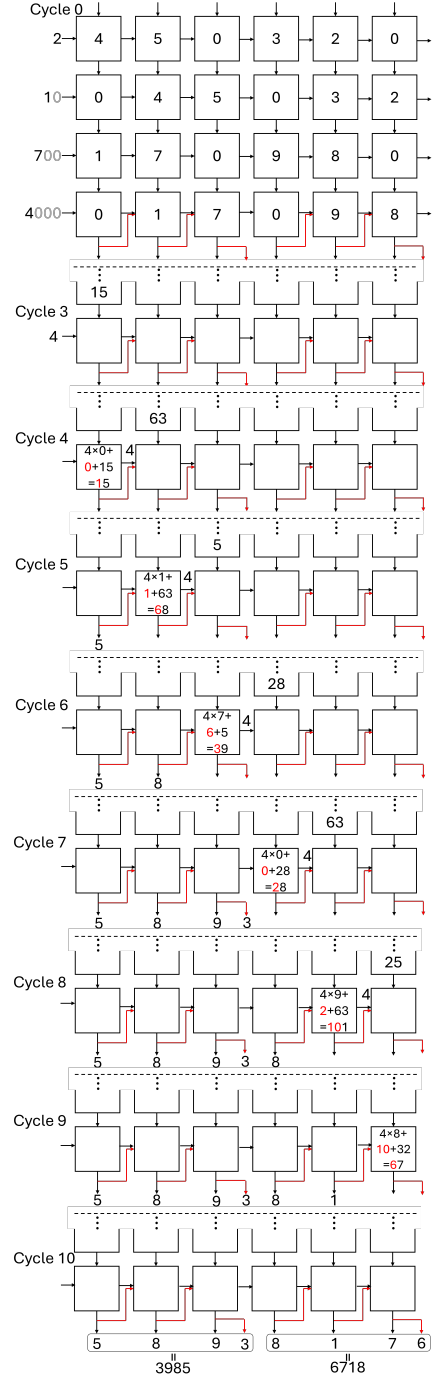}
    \caption{Running example on the in-sync execution of per-column vertical reduction needed for single-digit matrix multiplication and horizontal digit-level carry propagation for recomposing full-precision matrix multiplication output.}
    \label{fig:example}
\end{figure}



Figure~\ref{fig:example} presents a cycle-by-cycle illustration of reconstruction performed concurrently with the systolic wavefront of matrix multiplication as it propagates through the array. Initially, the right-hand matrix obtained from the decomposition in Fig.~\ref{fig:highgemm} is assumed to be preloaded into the array, while the left-hand matrix is streamed in row by row from the left. To emphasize only on the operation of reconstruction, Fig.~\ref{fig:example} considers only the first row of the decomposed left matrix from Fig.~\ref{fig:highgemm}. 

As the systolic array operates, the first row of the intermediate result matrix is progressively generated along the bottom of the array. The first outputs begin to appear in consecutive columns starting at cycle 3. Those elements effectively enter the RPEs of the last row. Each RPE retains the least significant byte for its column and forwards the most significant byte as a carry to the adjacent RPE on the right. This digit-level carry propagation continues until the full precision result has been reconstructed. The required number of digits is fixed at design time, with the appropriate configuration applied at runtime.



Please note that the intermediate results are streamed in \emph{least-significant-digit-first order}. This ordering is critical since it ensures that digit $k$ arrives only after digit $k-1$ has produced its carry on the previous column. 

\subsection{NTT-Specific Full-Precision Matrix Decomposition}

This section describes how full-precision matrix decomposition occurs specifically for NTT and simplified using the Basis-Aligned Transformation (BAT)~\cite{11408507}. The 4-step NTT factorizes an $N$-point transform into smaller transforms, computed as the product of an $N \times N$ input matrix $A$ and a twiddle-factor matrix $T$.
Suppose $A$ and $T$ contain $b$-bit full-precision integers. To execute this multiplication on the proposed systolic array, which only supports $w$-bit multipliers ($w < b$), both matrices are decomposed using the methodology from Section~\ref{sec:decomp}. Each entry is decomposed into
$$k=\left\lceil \frac{b}{w}\right\rceil$$
radix-$2^w$ digits. These digits correspond to positional weights $2^0, 2^w, \dots, 2^{(2k-2)w}$. This expansion structurally resizes matrix $A$ from $N \times N$ to $N \times kN$, and matrix $T$ from $N \times N$ to $kN \times (2k-1)N$.




Matrix $T$ contains only twiddle factors that depend on the transform parameters $(N,q)$ and are constant for a specific NTT instance. For this matrix, BAT suggests a way to reduce the size of this matrix taking into account that the contribution of each digit of decomposed matrix $T$ should be weighted modulo $q$. Thus, all digit weights that exceed $2^{kw}$ can be folded back to lower digits. Since matrix $T$ is known at design time and constant at runtime this reallocation of digits can be performed offline. 

The following toy example illustrates the offline reduction applied to twiddle factor digits. Consider the value $54$ (the first element of the right matrix in Fig.~\ref{fig:highgemm}) as one of the twiddle factors.  
Using base-10 digit decomposition with two-digit precision, the value is expanded into the following Toeplitz matrix:
\[
54 \rightarrow
\begin{bNiceMatrix}[first-row]
10^0 & 10^1 & 10^2 \\
4  & 5 & 0 \\
0  & 4 & 5 
\end{bNiceMatrix}
%
\]


Assuming that all computations are performed modulo $97$\, the contribution of digit 5 with weight $10^2$ can be reduced offline to $5\cdot 10^2 \bmod 97 = 15 = 1\cdot 10^1 + 5\cdot 10^0$. Distributing this result to the columns with weights $10^0$ and $10^1$, respectively, yields
\[
\begin{bNiceMatrix}[first-row]
10^0 & 10^1 & 10^2 \\
4  & 5 & 0 \\
0  & 4 & 5 
\end{bNiceMatrix}
\rightarrow\;
\begin{bNiceMatrix}[first-row]
10^0 & 10^1  \\
4  & 5  \\
0+\textcolor{red}{5}  & 4+\textcolor{red}{1}
\end{bNiceMatrix}
%
=
\begin{bmatrix}
4 & 5 \\
5 & 5
\end{bmatrix}.
\]
Twiddle-factor digit-folding is applied incrementally until all digits of the $T$ matrix fit in $w$ bits.
After applying offline this redistribution of twiddle factors as suggested by BAT~\cite{11408507} the dimension of decomposed matrix $T$ reduces to $k\, N\times k\, N$.  

\section{Experimental Evaluation}
\label{sec:results}


To evaluate the proposed approach, we quantify both the performance speedup achieved when executing FHE NTT kernels (for transform sizes ranging from $2^{12}$ to $2^{16}$) using the 4-step NTT algorithm, and the minimal hardware overhead introduced relative to a standard TPU. For performance measurements, we utilize the cycle-accurate SCALE-Sim simulator~\cite{11096402}, which models architectures similar to commercial AI accelerators. The simulated baseline features a $128\times 128$ systolic matrix unit (MXU) paired with a 128-lane vector processing unit (VPU)~\cite{9499913}. 
Each MXU integrates dedicated 8 MB on-chip input, weight, and output buffers with sufficient bandwidth to sustain peak systolic throughput. These buffers deliver one operand per processing element per cycle (e.g., 128 8-bit words/cycle for a 128×128 array), ensuring continuous utilization of the array.


We compare the proposed design with two state-of-the-art approaches: (a) TensorFHE~\cite{10071017} executes 
low-precision matrix multiplications on the MXU and the full-precision reconstruction is offloaded to the VPU; (b) CROSS~\cite{11408507} follows the same approach as TensorFHE for matrix multiplication and result reconstruction but follows a different decomposition of the original matrices to their low-precision counterparts. The decomposition of CROSS is similar to the one described in Section~\ref{sec:decomp}. 
In all cases, including the proposed approach, when matrix multiplication sizes exceeds the size of MXU, multiplication is performed in tiles~\cite{schonhage1971schnelle}.

Because FHE schemes dictate 32- to 64-bit modular arithmetic representing the most commonly used moduli in lattice-based cryptography and typical FHE libraries, we evaluate both configurations.

\subsection{Runtime performance comparisons}

\begin{table}[t]
\centering
\caption{Runtime comparison (in cycles) and relative speedup of the 4-step NTT algorithm between TensorFHE~\cite{10071017}, CROSS~\cite{11408507} and the proposed architecture for 32-bit and 64-bit full-precision inputs across varying transform sizes.}
\label{tab:results}
\begin{subtable}{\linewidth}
\centering
\label{tab:results32}
\setlength{\tabcolsep}{1pt}
\begin{tabular}{|c|ccc|ccc|ccc|}\hline
 \multirow{3}{*}{\shortstack{NTT \\ Size}} & \multicolumn{3}{c|}{TensorFHE~\cite{10071017}} & \multicolumn{3}{c|}{CROSS~\cite{11408507}} & \multicolumn{3}{c|}{Proposed}\\
 &  MXU& VPU & \multirow{2}{*}{Total}  & MXU & VPU & \multirow{2}{*}{Total} & MXU & VPU &\multirow{2}{*}{Total}  \\
 & cycles & cycles & & cycles & cycles & & cycles & cycles & \\\hhline{==========}
 $2^{12}$ & 14240 & 2176  & 16416  & 3566   & 2176  & 5742   & 3566  & 256  & 3822  \\
 $2^{13}$ & 15264 & 4288  & 19552  & 9174   & 4288  & 13462  & 9174  & 448  & 9622  \\
 $2^{14}$ & 16288 & 8576  & 24864  & 16318  & 8576  & 24894  & 16318 & 896  & 17214 \\
 $2^{15}$ & 42816 & 17024 & 59840  & 42846  & 17024 & 59328  & 42846 & 1664 & 44510 \\
 $2^{16}$ & 81632 & 34048 & 115680 & 81662  & 34048 & 115710 & 81662 & 3328 & 84990 \\\hline
\end{tabular}\\
\caption{32-bit data width}
\end{subtable}
\begin{subtable}{\linewidth}
\centering
\label{tab:results32}
\setlength{\tabcolsep}{1pt}
\begin{adjustbox}{width=\linewidth}
\begin{tabular}{|c|ccc|ccc|ccc|}\hline
 \multirow{3}{*}{\shortstack{NTT \\ Size}} & \multicolumn{3}{c|}{TensorFHE~\cite{10071017}} & \multicolumn{3}{c|}{CROSS~\cite{11408507}} & \multicolumn{3}{c|}{Proposed}\\
 &  MXU & VPU  & \multirow{2}{*}{Total}  & MXU & VPU  & \multirow{2}{*}{Total} & MXU & VPU   & \multirow{2}{*}{Total} \\
 & cycles & cycles  & & cycles & cycles  &  &  cycles & cycles &  \\\hhline{==========}
 $2^{12}$ & 56960  & 8320    & 65280  & 14270    & 8320    & 22590  & 14270  & 256  & 14526   \\
 $2^{13}$ & 61056  & 16576   & 77632  & 36702    & 16576   & 53278  & 36702  & 448  & 37150   \\
 $2^{14}$ & 65152  & 33152   & 98304  & 65278    & 33152   & 98430  & 65278  & 896  & 66174   \\
 $2^{15}$ & 171264 & 66176   & 237440 & 171390   & 66176   & 237566 & 171390 & 1664 & 173054  \\
 $2^{16}$ & 326528 & 132352  & 458880 & 326654   & 132352  & 459006 & 326654 & 3328 & 329982  \\\hline
\end{tabular}
\end{adjustbox}
\caption{64-bit data width}
\end{subtable}
\end{table}

Table~\ref{tab:results} summarizes the runtime comparison between TensorFHE, CROSS, and the proposed architecture for 32-bit and 64-bit full-precision inputs. For each NTT size, the table reports the total number of cycles required to execute a complete 4-step NTT. For all architectures, the execution time is divided into MXU cycles, corresponding to 8-bit digit matrix multiplications, and VPU cycles, corresponding to all operations performed outside the systolic array. In TensorFHE and CROSS, VPU cycles include the shift-add operations required to reconstruct full-precision results, as well as modular reductions, twiddle-factor multiplications, and matrix transpositions involved in the 4-step NTT. In contrast, for the proposed architecture, VPU cycles account only for modular reduction, since full-precision reconstruction is performed online during matrix multiplication by the RPEs within the systolic array.


Across both precision settings, the proposed architecture consistently reduces execution cycles, achieving speedups between $1.33\times$ and $4.49\times$. The results for 32-bit and 64-bit operands exhibit nearly identical trends and speedups.


The primary source of performance improvement is the elimination of VPU-based full-precision output reconstruction that involves shift-add operations that are not natively supported by typical MXUs. In TensorFHE~\cite{10071017} and CROSS~\cite{11408507}, a non-negligible portion of execution time is spent outside the systolic array combining partial sums and performing digit-level carry propagation. By integrating these operations directly into the MXU dataflow, the proposed architecture removes this overhead entirely.

The largest speedups are observed for smaller NTT sizes against TensorFHE~\cite{10071017}. For example, in the 32-bit configuration at size $2^{12}$, execution time is reduced from 16416 cycles to 3822 cycles, corresponding to a $4.30\times$ improvement. In this regime, VPU execution constitutes a significant fraction of total runtime, so eliminating it yields substantial gains.
As the NTT size increases, matrix multiplication increasingly dominates execution time. Consequently, the relative impact of removing VPU overhead decreases and speedups converge to  1.36$\times$ for the largest transform size $2^{16}$. 


\subsection{Physical synthesis comparisons}
\begin{figure}[t]
    \centering
    \includegraphics[width=0.55\linewidth]{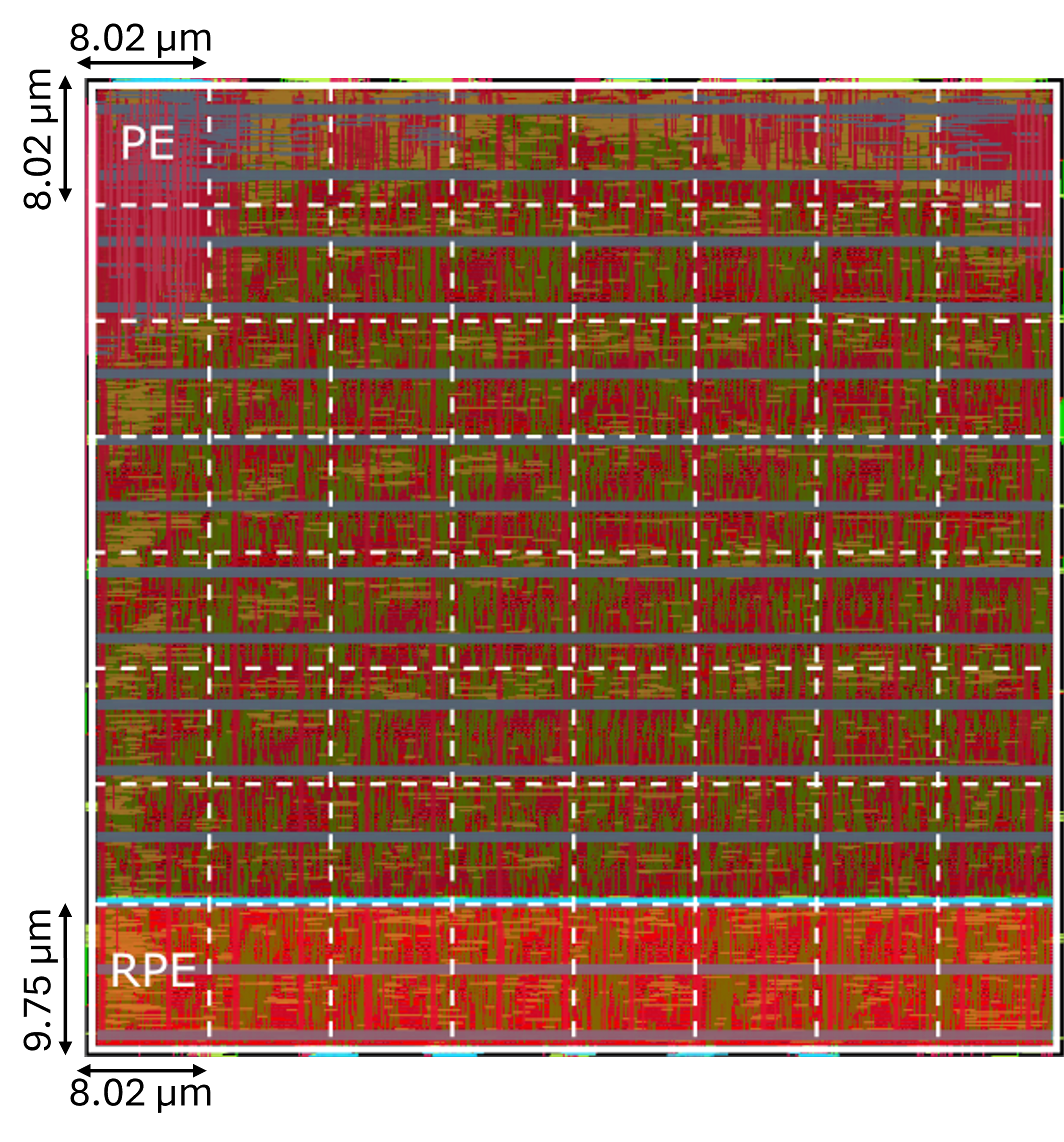}
    \caption{The physical layout of an $8\times 8$ instance of the proposed systolic array.}
    \label{fig:layout}
\end{figure}

To evaluate the hardware overhead introduced by the extra adder and multiplexing logic in the RPEs, which replace the final row of standard PEs in the baseline systolic array, we implemented both the proposed design and the baseline using the OpenROAD physical synthesis platform with the ASAP7 technology library targeting a clock frequency of 1 GHz that matches industry-grade TPUs~\cite{9499913}. As an example, Fig.~\ref{fig:layout} shows the physical layout of a small-scale $8\times 8$ instance of the proposed systolic array.

Table~\ref{tab:areapower} reports the post-layout area and power consumption for both architectures across multiple array sizes, alongside the relative overhead introduced by the additional full-precision output reconstruction logic. The results demonstrate that the proposed modifications incur minimal hardware penalties. Specifically, the area overhead decreases from 0.35\% for small arrays to just 0.09\% for larger configurations, while the power overhead drops from 0.69\% to 0.17\%. This scaling trend reflects the amortization of the final row's additional logic as the overall array size increases. 

Overall, the proposed approach reduces the total energy per NTT by significantly lowering latency with only negligible power overhead, demonstrating that native full-precision output reconstruction within the systolic array is both an effective performance optimization and a practical energy efficient hardware enhancement.

\begin{table}[t]
\centering
\caption{Area and power comparison between a conventional systolic array and the proposed enhanced design with RPEs in the last row, implemented at 7 nm using the OpenROAD physical synthesis platform.}
\label{tab:areapower}
\setlength{\tabcolsep}{3.5pt}
\begin{tabular}{|c|ccc|ccc|}\hline
\multirow{2}{*}{MXU Size} & \multicolumn{3}{c|}{Area (mm$^2$)} & \multicolumn{3}{c|}{Power (W)} \\
& baseline & proposed & overhead & baseline & proposed & overhead \\ \hhline{=======}
$8   \times   8$ & 0.0041 & 0.0042 & 2.79\% & 0.0254  & 0.0241  & 5.52\% \\
$16  \times  16$ & 0.0165 & 0.0167 & 1.39\% & 0.0989  & 0.0963  & 2.76\% \\
$32  \times  32$ & 0.0659 & 0.0664 & 0.70\% & 0.3903  & 0.3850  & 1.38\% \\
$64  \times  64$ & 0.263  & 0.264  & 0.35\% & 1.84    & 1.86    & 0.69\% \\
$128 \times 128$ & 1.054  & 1.056  & 0.17\% & 7.39    & 7.41    & 0.34\% \\
$256 \times 256$ & 4.217  & 4.221  & 0.09\% & 29.56   & 29.61   & 0.17\% \\ \hline
\end{tabular}
\end{table}
\section{Conclusion}
Executing FHE kernels, such as the NTT, on AI ASICs offers a widely available, energy-efficient alternative to custom hardware. However, a fundamental discrepancy exists: AI ASICs are tuned for low-precision arithmetic, whereas FHE requires high precision. Mapping high-precision workloads onto low-precision matrix engines using standard dataflows is possible but creates serialization bottlenecks during full-precision result reconstruction. To address this, we propose a novel multi-precision systolic array architecture that unifies low-precision matrix multiplication with in-situ high-precision reconstruction. Our design introduces a novel dataflow where matrix multiplication results flow vertically, perfectly synchronized with horizontal, digit-level carry propagation. This fused approach eliminates reconstruction bottlenecks, delivering significant execution speedups for NTT kernels with a hardware overhead of less than 1\%.

\bibliographystyle{IEEEtran}
\bibliography{refs}

\end{document}